\documentclass{appolb}
\usepackage{graphicx}
\usepackage{lineno}

\begin{document}

\title{Mixed harmonic charge dependent azimuthal correlations in Pb-Pb
  collisions at $\sqrt{s_{NN}}=2.76$ TeV measured with the ALICE
  experiment at the LHC}

  \author{Y.Hori for the ALICE Collaboration\\
    \address{
      {\it
	Center for Nuclear Study, Graduate School of Science, University
	of Tokyo
      }
    }
}
\maketitle

\begin{abstract}
Mixed harmonic charge dependent azimuthal correlations at mid-rapidity in
Pb-Pb collisions at $\sqrt{s_{NN}}=2.76$ TeV were measured with the ALICE detector
at the LHC. A clear charge dependence for a series of correlations is observed both via the
multi-particle cumulant and the event plane methods. 
Implications from these measurements for the possible
effects of local parity violation in QCD and for models
which incorporate azimuthal anisotropic flow and ``effective" local charge conservation on
the kinetic freeze-out surface are discussed. 

\end{abstract}
 
\section{Introduction}

The charge dependence of azimuthal correlations
between produced hadrons is an important probe
of QGP matter created in relativistic heavy-ion collisions.
It is in particular sensitive to the interplay between the
local charge conservation (LCC) induced correlations and azimuthally asymmetric radial expansion
of the collision system \cite{a05}.

Recently, it was argued that the charge dependent azimuthal
correlations can be also sensitive to the possible effect of local parity violation in QCD \cite{a02}.
Parity violation in QCD may happen as a result of the interaction between
produced quarks and topologically non-trivial gluonic field configurations.
In the presence of the strong magnetic field generated in a heavy-ion collision,
local parity violation may result in a separation of charges along the magnetic field
which points perpendicular to the reaction plane.
This phenomenon is called the Chiral Magnetic Effect (CME).

An important observable which was proposed as a sensitive probe of the
CME is the two particle correlation with respect to the reaction plane $\langle
{\rm cos}(\varphi_{\alpha}+\varphi_{\beta}-2\Psi_{RP})\rangle$
\cite{a03}, where the bracket denotes the average over all particles in all events
and the indices $\alpha$ and $\beta$ refer to the charge of the
particles.
$\varphi_{\alpha, \beta}$ is the azimuthal angle of the charged particles
and $\Psi_{RP}$ is the reaction plane angle.
In the presence of the CME, this correlation can be decomposed as
\begin{eqnarray}
  \langle {\rm cos} (\varphi_{\alpha}+\varphi_{\beta}-2\Psi_{RP}) \rangle
  &\sim& - \langle a_{\alpha} a_{\beta} \rangle + B_{in} - B_{out}, \nonumber
\end{eqnarray}
where $a_{\alpha,\beta}$ indicates the charge asymmetry due to the CME.
The average $\langle a_{\alpha} a_{\beta} \rangle$ is expected to be positive for the same
charge combination and negative for the opposite charge combination. $B_{in/out}$ denotes the
backgrounds when both two particles are in the in-plane/out-of-plane region.
Measurements by the STAR Collaboration revealed
non-zero charge dependent and independent parts of the correlation $\langle
{\rm cos}(\varphi_{\alpha}+\varphi_{\beta}-2\Psi_{RP})\rangle$, which are consistent
with qualitative expectations from the CME \cite{a04}.
However, a study \cite{a05} showed
that a significant part of the observed charge dependent part can be
described by the Blast Wave model incorporating effects of LCC on the
kinetic freeze-out surface.
Furthermore, the charge independent part may have 
non-zero contributions from directed flow fluctuations and effects
of momentum conservation \cite{a09, a10}.

Recently, the ALICE Collaboration released a paper \cite{a07}
where the correlation $\langle
{\rm cos}(\varphi_{\alpha}+\varphi_{\beta}-2\Psi_{RP})\rangle$ were measured at the
LHC energy.
In these proceedings, we extend the ALICE measurement with additional mixed harmonic charge dependent correlations, which may help to disentangle
the CME and LCC induced correlations [9].
We present the correlations $\Delta \langle {\rm cos} [n(\varphi_{\alpha} - \varphi_{\beta})] \rangle$
and $\Delta \langle {\rm cos} [\varphi_{\alpha} -
  (m+1)\varphi_{\beta}+m\Psi_{2}] \rangle$ where $n,m$ are integers
and $\Psi_{2}$ is an azimuthal angle of the 2nd order collision symmetry plane. 
Here $\Delta$ denotes the
difference between the same and opposite charge correlations.
In terms of the LCC, the charge dependent part of the correlation
 $\Delta \langle {\rm cos} [n(\varphi_{\alpha} - \varphi_{\beta})] \rangle$
measures moments of the azimuthal distribution between balancing charges.
The first moment ($n$ = 1) is connected to the inverse width of the
distribution of the balancing charges, which is sensitive to the radial expansion of
the system.
Similarly, the charge dependent part of the correlation 
$\Delta \langle {\rm cos} [\varphi_{\alpha} -  (m+1)\varphi_{\beta}+m\Psi_{2}] \rangle$
measures a modulation of charge balancing width due to the $m$-harmonic
anisotropic flow relative to the 2nd collision order symmetry plane.

\section{Analysis details}
A sample of about 13 M minimum bias Pb-Pb collisions at
$\sqrt{s_{NN}} = 2.76$ TeV
collected by the ALICE detector during the 2010 LHC run was analyzed.
A description of the ALICE detector and details about collision triggers
and centrality determination
can be found in \cite{a07, ALICE-PPR}.
A Time Projection Chamber (TPC) is used to reconstruct charged particles
in the kinematic range $|\eta| < 0.8$ and $p_{\rm T} > 0.2$ GeV/$c$.
Correlations with respect to the symmetry plane
were measured
using the event plane and multi-particle cumulant methods.
In the event plane method, the symmetry planes were estimated
from azimuthal distributions of hits in two forward scintillator
counters (VZERO)
which cover the pseudo-rapidity range $-3.7 <  \eta < -1.7$ and $2.8 <
\eta  < 5.1$,
and two Forward Multiplicity Detectors (FMD) located at $1.7 < \eta <
5.1$ and $-3.4 < \eta < -1.7$.
In the multi-particle cumulant method, the correlations with respect
to the symmetry plane are evaluated from the azimuthal angle correlations of charged
particles reconstructed by the TPC.
Although the dominant systematic errors come from the
  event plane determinations,
  we observed good agreement between results from the different methods.
\section{Results}
The centrality dependence of the correlations $\langle
{\rm cos}(\varphi_{\alpha}-\varphi_{\beta})\rangle$ and $\langle
{\rm cos}(\varphi_{\alpha}+\varphi_{\beta}-2\Psi_{RP})\rangle$ for the same
and opposite charge combinations measured for Pb-Pb collisions
at $\sqrt{s_{NN}} = 2.76$ TeV was reported by ALICE in [7].
For the correlation $\langle
{\rm cos}(\varphi_{\alpha}+\varphi_{\beta}-2\Psi_{RP})\rangle$, ALICE observed that
the same charge correlation is non-zero and negative, while the opposite charge
correlation has a significantly smaller magnitude and is positive for peripheral collisions.
ALICE also showed that there is little collision energy dependence when comparing 
results to that at the top RHIC energy.
Even though some of the features of the observed correlations are in qualitative
agreement with the expectation from the CME,
origins of both charge dependent and independent parts are still not clear since they are
also sensitive to
many other parity-conserving physics mechanisms.
To study the physics backgrounds for the CME search, ALICE has measured the two
particle correlation $\langle
{\rm cos}(\varphi_{\alpha}-\varphi_{\beta})\rangle$, which also shows strong charge dependence but its
correlation strength is significantly different from what was measured
by the STAR Collaboration
at lower collision energy. 
This correlation may have a contribution of the CME,
\begin{eqnarray}
 \langle {\rm cos} (\varphi_{\alpha} - \varphi_{\beta}) \rangle &\sim& \langle
 a_{\alpha} a_{\beta} \rangle + B_{in} + B_{out}, \nonumber 
\end{eqnarray}
but its measurement is expected to be dominated by the large
background correlations $B_{in} + B_{out}$, and in particular
by those unrelated to the reaction plane orientation (nonflow).
\begin{figure}[hh]
  \begin{tabular}{cc}
    \begin{minipage}{0.5\hsize}
      \centering
      \includegraphics[width=6cm]{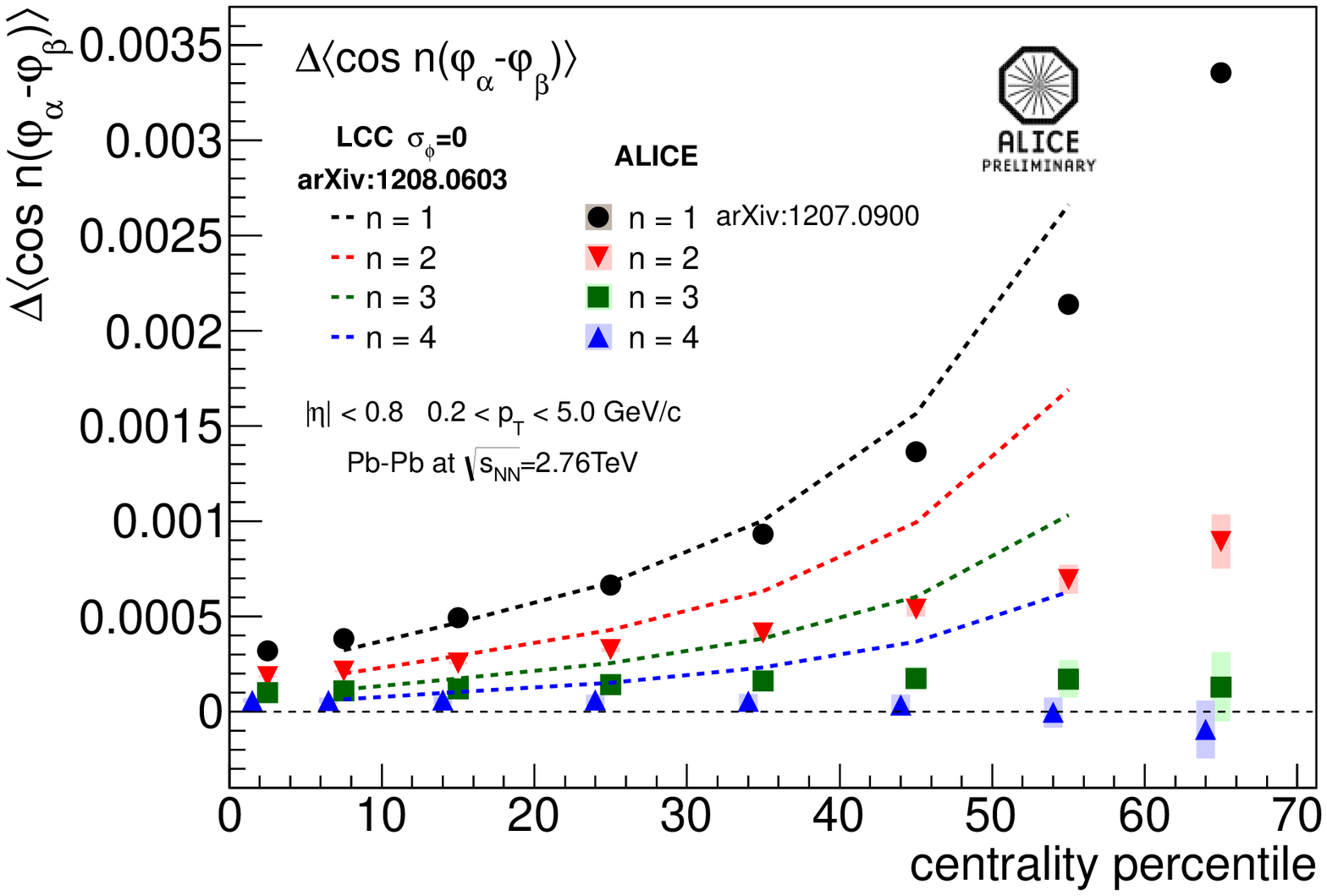}
    \end{minipage}
    \begin{minipage}{0.5\hsize}
      \centering
      \includegraphics[width=6cm]{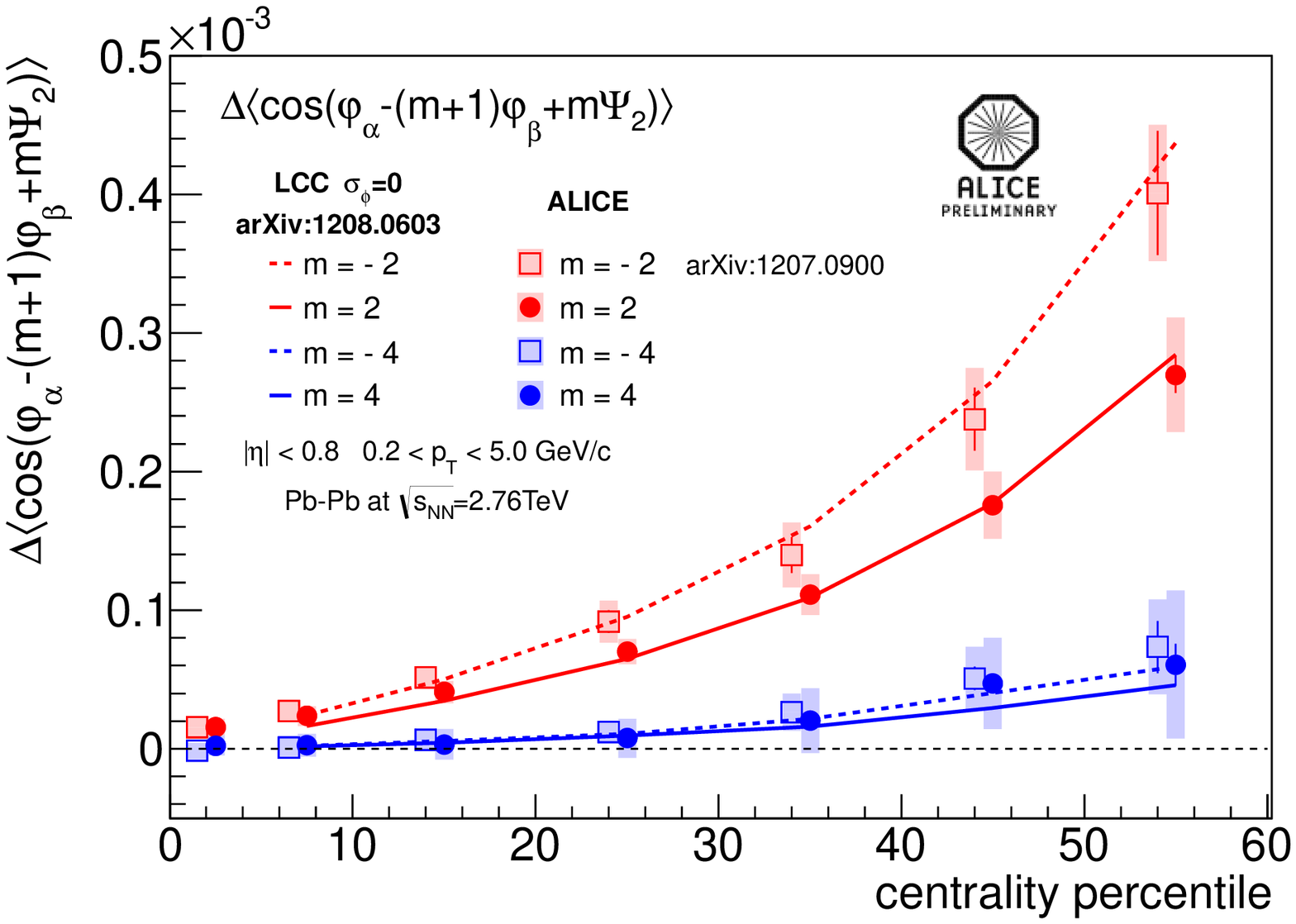}
    \end{minipage}
  \end{tabular}
  \caption{Centrality dependence of the charge dependent part of the correlation (left) $\Delta \langle
{\rm cos} [n(\varphi_{\alpha} - \varphi_{\beta})] \rangle$ and (right) $\Delta \langle {\rm
  cos} [\varphi_{\alpha}-(m+1)\varphi_{\beta}+m\Psi_{2}] \rangle$
in comparison with the Blast Wave model incorporating effects of LCC.}
\end{figure}

As shown in Fig. 1, measurements were extended to a set of charge
dependent correlations, which help to better constrain the possible physical contributions
to the previously measured correlation $\langle {\rm
  cos}(\varphi_{\alpha}+\varphi_{\beta} - 2\Psi_{RP})\rangle$.
Blast Wave parameters of the LCC model used in Fig. 1 are tuned on the measured
hadron spectra and the anisotropic flow at the LHC.
The large charge dependent part of the correlation $\langle
{\rm cos} (\varphi_{\alpha} - \varphi_{\beta}) \rangle$ can be
reproduced well by this LCC model while it fails to describe the
higher ($n>$1) harmonic correlations.
The right plot of Fig. 1 shows the charge dependent parts of the correlations
 $\langle {\rm
  cos} [\varphi_{\alpha}-(m+1)\varphi_{\beta}+m\Psi_{2}] \rangle$
in comparison with LCC model calculations.
Partial agreement between these measured correlations and the LCC
 model indicates that
the ``effective" LCC is indeed realized on the kinetic freeze-out surface. 
Therefore, the observed charge dependent part of the correlation $\langle {\rm
  cos}(\varphi_{\alpha}+\varphi_{\beta} - 2\Psi_{RP})\rangle$ ($m=-2$
 in Fig. 1 (right)) can be interpreted mostly as the LCC induced correlation.
More studies are needed to quantify the actual contributions
from the effects of LCC in the CME studies.
ALICE also measured
a charge dependent part of the two particle correlation with respect to
 the 3rd and 4th order collision symmetry planes, which also may help in disentangling effects from LCC and
CME  \cite{Y.Hori-QM2012-proceedings,
  S.Voloshin-QM2012-proceedings}.

It was suggested in [5] that non zero mixed harmonic correlations
associated with the possible directed flow $v_{1}$ may
be generated by the initial energy density fluctuations and hydrodynamic
expansion of the system created in a heavy-ion collision.
This effect may contribute a charge independent backgrounds for the
CME search.
Figure 2 shows the measured charge independent part of the correlations $\langle {\rm cos}(\varphi_{\alpha}-\varphi_{\beta})
\rangle$ (left) and $\langle {\rm
  cos}[\varphi_{\alpha}-(m+1)\varphi_{\beta}+m\Psi_{2}]\rangle$ (right).
An estimate of the charge independent correlation
with HIJING event generator in Fig. 2 (left) indicates
large non-flow contribution to the charge independent part
of the correlations $\langle {\rm cos}(\varphi_{\alpha}-\varphi_{\beta})
\rangle$ \cite{HIJING}.
At the same time, a rough agreement between the data
and AMPT model \cite{AMPT} supports the interpretation
of the charge independent correlations with respect to the 2nd order
collision symmetry plane
in terms of the event-by-event initial energy fluctuations.
Further studies, in particular the comparison with the differential
dependencies shown in Fig. 3 and in \cite{Ante}, are needed to make firm conclusions.

\begin{figure}[hh]
  \begin{tabular}{cc}
    \begin{minipage}{0.5\hsize}
      \centering
      \includegraphics[width=6cm]{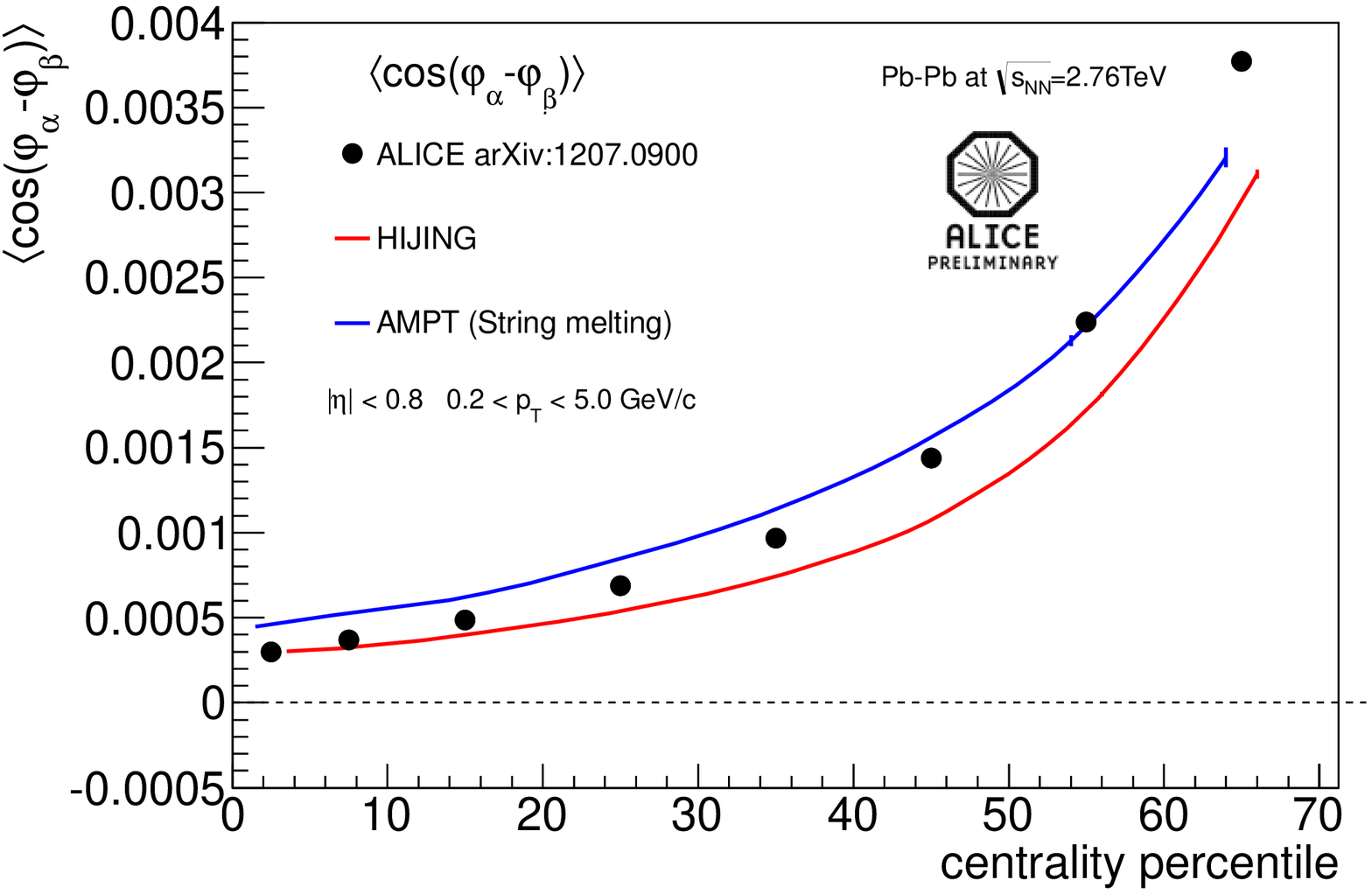}
    \end{minipage}
    \begin{minipage}{0.5\hsize}
      \centering
      \includegraphics[width=6cm]{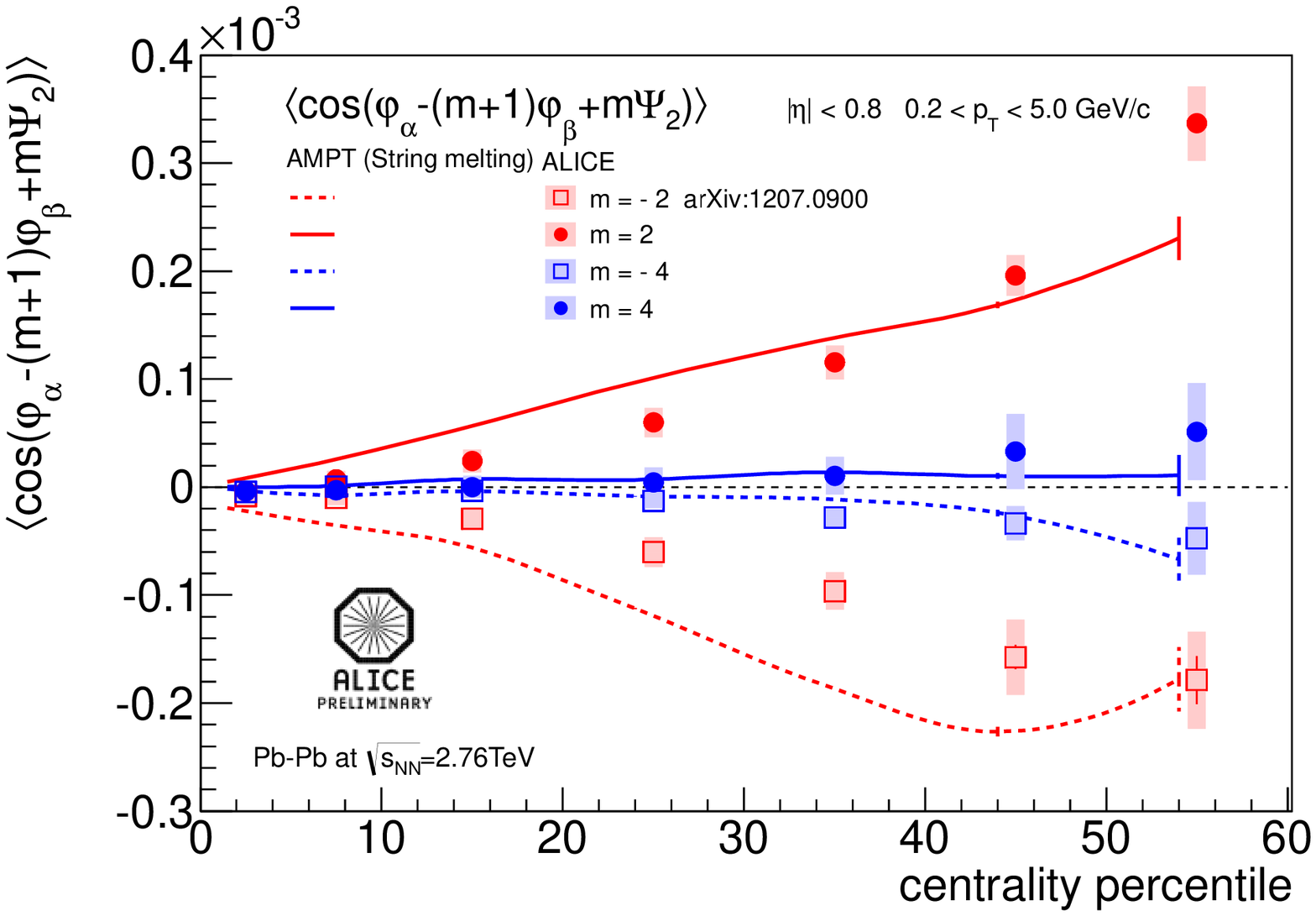}
    \end{minipage}
  \end{tabular}
  \caption{Centrality dependence of the charge independent part of the
    correlation (left) $\langle
    {\rm cos} (\varphi_{\alpha} - \varphi_{\beta}) \rangle$ and (right) $\langle
    {\rm cos} [\varphi_{\alpha} - (m+1)\varphi_{\beta} + m\Psi_{2}] \rangle$ measured by the
    ALICE in comparison with HIJING and AMPT with
    the {\it string melting} configuration \cite{HIJING, AMPT}.}
\end{figure}

Differential studies of charged dependent correlations
vs. pair transverse momentum and pseudo-rapidity
provide further constraints on models.
Figure 3 shows the pair differential dependencies of the correlation
$\langle {\rm cos} (\varphi_{\alpha}-3\varphi_{\beta}+2\Psi_{RP}) \rangle$.
Similarly to the results reported in [7] for the correlation
$\langle {\rm cos} (\varphi_{\alpha}+\varphi_{\beta}-2\Psi_{RP}) \rangle$,
we observe that the correlation is localized within about one unit of
rapidity
(or may even change sign as a function of $\Delta \eta$)
and extends up to the higher $p_{\rm T}$ of the pair.

\begin{figure}[hh]
  \centering
  \includegraphics[width=10cm]{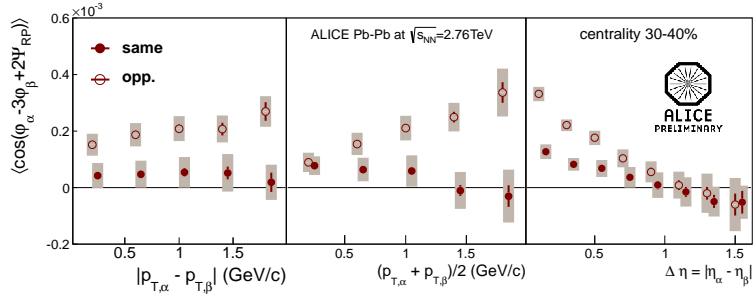}
  \caption{The pair differential correlation 
    $\langle {\rm
      cos} (\varphi_{\alpha}-3\varphi_{\beta}+2\Psi_{RP}) \rangle$ as a function
  of (left) the transverse momentum difference
  $|p_{{\rm T},\alpha}-p_{{\rm T},\beta}|$,
(center) the average transverse momentum $(p_{{\rm T},\alpha}+p_{{\rm T},\beta})/2$,
  (right) the rapidity separation $\Delta \eta =
  |\eta_{\alpha}-\eta_{\beta}|$ of the charged particle pair. }
\end{figure}
\section{Summary}
Charge dependent azimuthal correlations 
in Pb-Pb collisions at $\sqrt{s_{NN}}=2.76$ TeV were measured by the ALICE Collaboration.
A significant non-zero correlation 
$\langle {\rm cos}(\varphi_{\alpha}+\varphi_{\beta}-2\Psi_{RP})
\rangle$ was observed, which
was originally proposed as an observable sensitive to the CME and thus to effects from local parity violation in QCD.
The experimental analysis was extended to the higher moments of the
two particle azimuthal correlations
$\langle {\rm cos} [n(\varphi_{\alpha}-\varphi_{\beta})] \rangle$ for $n=1-4$ and to the mixed harmonic
charge dependent azimuthal correlations with respect to the 2nd order collision symmetry plane
(e.g. $\langle {\rm cos}(\varphi_{\alpha}-3\varphi_{\beta}+2\Psi_{2}) \rangle$).
These new results provide an important experimental input which is
relevant to CME, ``effective" LCC on the kinetic freeze-out, and directed flow fluctuations.






\bibliographystyle{elsarticle-num}
\bibliography{<your-bib-database>}











\end{document}